\PassOptionsToPackage{dvipsnames}{xcolor}
\PassOptionsToPackage{pagebackref,draft=false}{hyperref}
\documentclass{article}

\usepackage{geometry}
\usepackage{amsmath,amsthm,amsfonts}
\usepackage{colonequals}
\usepackage{todonotes}
\usepackage{mathtools}
\usepackage{tikz,tikz-cd}
\usepackage{tikzit}				

\tikzstyle{morphism}=[fill=white, draw=black, shape=rectangle]
\tikzstyle{medium box}=[fill=white, draw=black, shape=rectangle, minimum width=0.8cm, minimum height=0.9cm]
\tikzstyle{large morphism}=[fill=white, draw=black, shape=rectangle, minimum width=1.7cm, minimum height=1cm]
\tikzstyle{bn}=[fill=black, draw=black, shape=circle, inner sep=1.5pt]
\tikzstyle{state}=[fill=white, draw=black, regular polygon, regular polygon sides=3, minimum width=0.8cm, shape border rotate=180, inner sep=0pt]
\tikzstyle{medium state}=[fill=white, draw=black, regular polygon, regular polygon sides=3, minimum width=1.3cm, inner sep=0pt, shape border rotate=180]
\tikzstyle{large state}=[fill=white, draw=black, regular polygon, regular polygon sides=3, minimum width=2.2cm, shape border rotate=180, inner sep=0pt]
\tikzstyle{wn}=[fill=white, draw=black, shape=circle, inner sep=1.5pt]
\tikzstyle{wide state}=[fill=white, draw=black, shape=isosceles triangle, minimum width=0.8cm, shape border rotate=270, inner sep=1.4pt, minimum height=0.5cm, isosceles triangle apex angle=80]
\tikzstyle{evalold}=[fill=white, draw=black, shape=isosceles triangle, minimum width=1.4cm, shape border rotate=90, inner sep=1.4pt, minimum height=0.4cm, isosceles triangle apex angle=110]
\tikzstyle{eval}=[fill=white, draw=black, shape=rectangle, minimum width=1.4cm, minimum height=0.55cm, inner sep=1.4pt, font={$\eval$}]

\tikzstyle{arrow}=[->]
\tikzstyle{dashed box}=[-, dashed]
\tikzstyle{mapsto}=[{|->}]
\tikzstyle{double wire}=[-, double]
\tikzstyle{protected}=[-, preaction={{ultra thick,white,draw}}]
\tikzstyle{ambient fill}=[-, draw=none, fill={rgb,255: red,245; green,220; blue,255}, tikzit draw={rgb,255: red,210; green,130; blue,255}]
	\usetikzlibrary{external}
\usepackage{enumitem}
	\setlist[itemize]{itemsep=0pt,topsep=4pt}
	\setlist[enumerate]{itemsep=0pt,topsep=4pt}
	\setlist[enumerate,1]{label=(\roman*)}  
	\setlist[enumerate,2]{label=(\alph*)}  
\usepackage{tikz-cd}
\usepackage{amssymb,xparse}		
\usepackage{scalerel}
\usepackage{authblk}
\usepackage{bm}				
\usepackage{centernot}
\usepackage{doi} 
\usepackage{nicefrac}
\usepackage{nameref} 
\usepackage[numbers,sort]{natbib}
\usepackage[verbose=true]{microtype} 
\usepackage[parfill]{parskip}		
\usepackage{makecell} 
\usepackage{array}

\allowdisplaybreaks

\definecolor{myurlcolor}{rgb}{0,0,0.3}
\definecolor{mycitecolor}{rgb}{0,0.3,0}
\definecolor{myrefcolor}{rgb}{0.3,0,0}
\usepackage[pagebackref,draft=false]{hyperref}
\hypersetup{colorlinks,
	linkcolor=myrefcolor,
	citecolor=mycitecolor,
urlcolor=myurlcolor}

\usepackage[noabbrev,capitalize]{cleveref}	

\definecolor{fillcolor}{rgb}{0.75,0.75,0.75}

\newtheorem*{theorem*}{Theorem}

\theoremstyle{definition}

\numberwithin{equation}{section}

\let\originalleft\left
\let\originalright\right
\renewcommand{\left}{\mathopen{}\mathclose\bgroup\originalleft}
\renewcommand{\right}{\aftergroup\egroup\originalright}

\newcommand{\eval}{\mathrm{eval}} 

\newcommand{\ph}{\mathord{\rule[-0.05em]{0.6em}{0.05em}}}				

\newcommand{\cat}[1]{{\mathsf{#1}}}

\newcommand{\id}{\mathrm{id}} 		

\tikzset{pullback/.style={minimum size=1.2ex,path picture={	
			\draw[opacity=1,black,-,#1] (-0.5ex,-0.5ex) -- (0.5ex,-0.5ex) -- (0.5ex,0.5ex);%
}}}

\newcommand{\cop}{\mathrm{copy}}
\newcommand{\discard}{\mathrm{del}}
\definecolor{parametrized}{RGB}{15,0,150}

\newcommand{\as}[1]{
	\def\relstate{#1}%
	\ifx\relstate\empty
		\text{a.s.}%
	\else
		{#1\text{-a.s.}}%
	\fi
}

\providecommand{\given}{}			
\newcommand{\SetSymbol}[1][]{%
	\nonscript\;\,#1\vert
	\allowbreak
	\nonscript\;\,
	\mathopen{}
}
\DeclarePairedDelimiterX{\Set}[1]{\{}{\}}{%
	\renewcommand{\given}{\SetSymbol[\delimsize]}
	#1
}
\makeatletter
\let\oldSet\Set
\def\Set{\@ifstar{\oldSet}{\oldSet*}}
\makeatother

\DeclarePairedDelimiterX{\Family}[1]{(}{)}{%
	\renewcommand{\given}{\SetSymbol[\delimsize]} 
	#1
}
\makeatletter
\let\oldFamily\Family
\def\Family{\@ifstar{\oldFamily}{\oldFamily*}}

\newsavebox{\numbox}%
\newsavebox{\slashbox}%
\newsavebox{\denbox}%
\newlength{\slashlength}%
\newlength{\faktorscale}%
\DeclareDocumentCommand{\newfaktor}{m O{0.35} m O{-0.35}}{
	\savebox{\numbox}{\ensuremath{#1}}
	\savebox{\slashbox}{\ensuremath{\diagup}}
	\savebox{\denbox}{\ensuremath{#3}}
	\setlength{\faktorscale}{0.5\ht\numbox+0.5\ht\denbox}%
	\setlength{\slashlength}{2pt+0.8\faktorscale+#2\faktorscale-#4\faktorscale}%
	\raisebox{#2\ht\slashbox}{\usebox{\numbox}}
	\mkern-2mu%
	\rotatebox{-30}{\rule[#4\ht\denbox]{0.4pt}{\slashlength}}
	\mkern9mu%
	\hspace{-0.44\slashlength}%
	\raisebox{#4\ht\denbox}{\usebox{\denbox}}
}
\DeclareDocumentCommand{\linefaktor}{m O{0.08} m O{-0.08}}{
	\savebox{\numbox}{\ensuremath{#1}}
	\savebox{\slashbox}{\ensuremath{\diagup}}
	\savebox{\denbox}{\ensuremath{#3}}
	\setlength{\faktorscale}{0.5\ht\numbox+0.5\ht\denbox}%
	\setlength{\slashlength}{0.2\faktorscale+0.8\baselineskip}%
	\raisebox{#2\ht\slashbox}{\usebox{\numbox}}
	\mkern-1mu%
	\raisebox{-0.8pt}{%
		\rotatebox{-25}{\rule[#4\ht\denbox]{0.4pt}{\slashlength}} 
	}%
	\mkern-1mu%
	\hspace{-0.25\slashlength}%
	\raisebox{#4\ht\denbox}{\usebox{\denbox}}
}

\newcommand{\newterm}[1]{\textbf{#1}}

\title{An Invitation to Universality in Physics, \\ Computer Science, and Beyond}

\author[1]{Tom{\'a}{\v{s}} Gonda}
\author[1]{Gemma De les Coves}

\affil[1]{Institute for Theoretical Physics, University of Innsbruck, Austria}

\begin{document}

\maketitle

\begin{abstract}
	A universal Turing machine is a powerful concept{\,---\,}a single device can compute any function that is computable.
	A universal spin model, similarly, is a class of physical systems whose low energy behavior simulates that of any spin system. 
	Our categorical framework for universality \cite{gonda2023framework} captures these and other examples of universality as instances.
	In this article, we present an accessible account thereof with a focus on its basic ingredients and ways to use it.
	Specifically, we show how to identify necessary conditions for universality, compare types of universality within each instance, and establish that universality and negation give rise to unreachability (such as uncomputability).
\end{abstract}

\section{Introduction}
\label{sec:Introduction}

	One of the most influential models of computation is that of Turing machines.
	Namely, a function is computable if it can be read off as the input-output \emph{behavior} of some Turing machine, 
	which constitutes a \emph{particular solution} to the information-theoretic \emph{problem} of computing the function.
	Naively, one may expect that building tailor-made Turing machines would be necessary to solve different problems.
	However, the existence of a universal Turing machine (a \emph{universal solution}) invites a wholly new perspective: 
	One may invest in the development of a universal machine and use it to tackle any information-theoretic problem, i.e.\ to compute any function of interest. 
	This is possible because, by definition, a universal Turing machine can \emph{simulate} any other Turing machine so long as it is provided with a \emph{program} carrying the requisite instructions.
	
	Spin systems are versatile toy models of complex systems \cite{So00,Th18}, which also exhibit universality \cite{De16b,Re23}. 
	More concretely, a spin system (the \emph{particular solution}) is a composite system with local interactions and an energy function.
	Individual degrees of freedom are called spins and each interaction contributes to the global energy additively by a real number that depends on the spin configuration. 
	The behavior of a spin system (the network-theoretic \emph{problem}) is a real-valued function of its spin configurations. 
Some families of spin systems, such as the two-dimensional (2D) Ising model with fields, are able to \emph{simulate} any other spin system. 
	In this sense, the 2D Ising model with fields is a universal spin model (the \emph{universal solution}).
	`Programming' it amounts to fixing the values of certain spins and coupling strengths of a suitably chosen 2D Ising spin system.
	
	In both situations, universal solutions can be programmed to solve any problem of interest; in other words, the universal solution can simulate any particular solution. 
Instead of building particular solutions, one may thus program universal ones. 
	Besides the pragmatic significance, universality is also of conceptual importance.
	In the case of computability, for example, it allows the study of algorithms with respect to a universal Turing machine, which leads to notions such as algorithmic complexity, algorithmic randomness, and algorithmic probability \cite{Li2008}.

	The main goal of our recent article titled ``A framework for universality in physics, computer science, and beyond'' \cite{gonda2023framework} is the development of a general theory of universal solutions.
	It elucidates how the two examples above are related and sets up a framework for investigations of novel instances of universality. 
	For example:
	\begin{enumerate}
		\item We prove a no-go theorem, which implies that a universal spin model cannot be finite (\cref{sec:no-go}).
		\item We distinguish the universality of Turing machines from that of spin models via the concept of a `singleton simulator' (\cref{sec:parsimony}).
		\item We show that the universality of Turing machines is non-trivial in a precise sense (\cref{sec:parsimony}).
		\item We generalize the connection between universality and uncomputability (\cref{sec:unreachability}).
	\end{enumerate}
	The purpose of this article is to summarize this framework with a focus on an accessible presentation.
We aim to provide an overview of its main ideas (\cref{sec:Simulators} and \cref{sec:universality}) as well as the four possible uses mentioned above.
	We refer to the longer paper \cite{gonda2023framework} throughout for details and technical aspects of the framework.
	
\section{Simulators}
\label{sec:Simulators}
	
	Both of our key examples of universal solutions can be viewed as collections of particular solutions.
	Specifically, a universal Turing machine is a collection with only one element, while a universal spin model usually contains infinitely many spin systems.
	We describe this collection by a map $s_T \colon P \to T$ called \newterm{compiler}, where:
	\begin{itemize}
		\item $T$ refers to the set of all particular solutions, also termed \newterm{targets}, e.g.\ $T$ is the set of all Turing machines when we study the universality of Turing machines.
		\item $P$ is a set of so-called \textbf{programs}, e.g.\ all finite strings.
		\item The image of $P$ under $s_T$ is a subset of particular solutions, e.g.\ a set consisting of a single universal Turing machine, in which case the compiler is a constant map.
	\end{itemize} 
On the other hand, the behavior of particular solutions is modeled by a function $\eval \colon T \otimes C \to B$ called \newterm{evaluation},\footnotemark{} where:
	\begin{itemize}
		\item $C$ is a set of so-called \newterm{contexts}, e.g.\ strings representing the possible states of the \emph{input} tape of a Turing machine.
		\item $B$ consists of (static) \textbf{behaviors}, e.g.\ strings representing the possible states of the \emph{output} tape of a Turing machine.
		\item For each target $t \in T$, the function $\eval(t, \ph) \colon C \to B$ represents the behavior of $t$, e.g.\ the (computable) function computed by the Turing machine $t$.
	\end{itemize}
	\footnotetext{The tensor product symbol $\otimes$ is used to denote parallel composition of objects.
	If $T$ and $C$ are objects of a category of sets and relations, for example, then $T \otimes C$ is the Cartesian product of sets $T$ and $C$. More generally, $\otimes$ is the monoidal product of a symmetric monoidal category.}%
	In order for a universal solution to simulate a particular solution, its behavior may be modified{\,---\,}in \cref{sec:Introduction} we refer to this procedure as the `programming' of the universal solution.
	Formally, it is given by a map $s_C \colon P \otimes C \to C$ called \newterm{context reduction}, where:
	\begin{itemize}
		\item For each program $p \in P$, the map $s_C(p, \ph) \colon C \to C$ specifies a pre-processing of the context for the target $s_T(p) \in T$, e.g.\ the function that takes an arbitrary string on the input tape and prepends an encoding of the string $p$ to it.
	\end{itemize}
The compiler and context reduction are combined into a map $s \colon P \otimes C \to T \otimes C$ called \newterm{simulator}, which models how particular solutions are simulated. 
	It can be depicted by the diagram on the left, while the right diagram can be thought of as the behavior of the simulator.
	\begin{equation}\label{eq:simulator}
		\begin{tikzpicture}
	\begin{pgfonlayer}{nodelayer}
		\node [style=morphism] (12) at (-7.5, 0.75) {$s_T$};
		\node [style=morphism] (13) at (-5, 0.75) {$\;\; s_C \;\;$};
		\node [style=none] (14) at (-7.5, 2) {};
		\node [style=none] (15) at (-7.5, 2.5) {$T$};
		\node [style=none] (16) at (-5, 2) {};
		\node [style=none] (17) at (-5, 2.5) {$C$};
		\node [style=none] (18) at (-5.5, 0.5) {};
		\node [style=none] (19) at (-4.5, 0.75) {};
		\node [style=bn] (20) at (-6.5, -0.5) {};
		\node [style=none] (21) at (-6.5, -1.5) {};
		\node [style=none] (22) at (-6.5, -2) {$P$};
		\node [style=none] (23) at (-4.5, -1.5) {};
		\node [style=none] (24) at (-4.5, -2) {$C$};
		\node [style=none] (25) at (-7.5, 0.5) {};
		\node [style=morphism] (27) at (4.75, -0.75) {$ \quad \; s \quad \;$};
		\node [style=none] (28) at (4, -1.75) {};
		\node [style=none] (29) at (4, -2.25) {$P$};
		\node [style=none] (30) at (5.5, -1.75) {};
		\node [style=none] (31) at (5.5, -2.25) {$C$};
		\node [style=none] (32) at (4, 1.25) {};
		\node [style=none] (34) at (5.5, 1.25) {};
		\node [style=none] (36) at (5.5, -0.75) {};
		\node [style=none] (37) at (4, -0.75) {};
		\node [style=eval] (38) at (4.75, 1.25) {};
		\node [style=none] (39) at (4.75, 2.25) {};
		\node [style=none] (40) at (4.75, 2.75) {$B$};
		\node [style=none] (41) at (-5.5, 0.75) {};
		\node [style=none] (42) at (-5.5, 0.5) {};
		\node [style=none] (43) at (3.5, 0.25) {$T$};
		\node [style=none] (44) at (6, 0.25) {$C$};
	\end{pgfonlayer}
	\begin{pgfonlayer}{edgelayer}
		\draw (12) to (14.center);
		\draw (13) to (16.center);
		\draw [in=-90, out=15] (20) to (18.center);
		\draw (23.center) to (19.center);
		\draw (21.center) to (20);
		\draw [in=-90, out=165] (20) to (25.center);
		\draw (30.center) to (36.center);
		\draw (36.center) to (34.center);
		\draw (28.center) to (37.center);
		\draw (37.center) to (32.center);
		\draw (38) to (39.center);
		\draw (25.center) to (12);
		\draw (42.center) to (41.center);
	\end{pgfonlayer}
\end{tikzpicture}
	\end{equation}
	Note that the program $p \in P$ used to generate a particular solution via the compiler is also used to update the context via the context reduction.
	This is ensured by the copying given by $\cop(p) \coloneqq (p,p) \in P \otimes P$ and depicted by the left diagram. 
	\begin{equation}\label{eq:copy_delete}
		\begin{tikzpicture}
	\begin{pgfonlayer}{nodelayer}
		\node [style=bn] (0) at (-5.5, 0) {};
		\node [style=none] (1) at (-6.5, 1) {};
		\node [style=none] (2) at (-4.5, 1) {};
		\node [style=none] (3) at (-5.5, -0.75) {};
		\node [style=none] (4) at (4.5, -0.5) {};
		\node [style=bn] (5) at (4.5, 0.75) {};
		\node [style=none] (6) at (-5.5, -1.25) {$P$};
		\node [style=none] (7) at (-6.5, 1.5) {$P$};
		\node [style=none] (8) at (-4.5, 1.5) {$P$};
		\node [style=none] (9) at (4.5, -1) {$P$};
		\node [style=none] (10) at (-8, 0) {$=$};
		\node [style=none] (11) at (-9.5, 0) {$\cop$};
		\node [style=none] (12) at (3, 0) {$=$};
		\node [style=none] (13) at (1.5, 0) {$\discard$};
	\end{pgfonlayer}
	\begin{pgfonlayer}{edgelayer}
		\draw [in=-90, out=165] (0) to (1.center);
		\draw [in=-90, out=15] (0) to (2.center);
		\draw (3.center) to (0);
		\draw (4.center) to (5);
	\end{pgfonlayer}
\end{tikzpicture}
	\end{equation}
	
	In the abstract version of our framework \cite[Definition 2.31]{gonda2023framework}, all sets above\footnotemark{} are generalized to objects in a symmetric monoidal category, and maps such as the compiler and the context reduction are morphisms thereof.
	\footnotetext{This is true for the specific case of a target--context category with \emph{intrinsic} behaviors \cite[Definition 2.31]{gonda2023framework}.
	In the case of a target--context category with behaviors \cite[Definition 2.26]{gonda2023framework}, the set $B$ of behaviors is not an object in the ambient category \cite[Section 2.2]{gonda2023framework}, but an object in the category of sets and relations.}%
	In addition, this category, referred to as the ambient category, is required to possess copying and deletion morphisms for every object, such as $\cop \colon P \to P \otimes P$ and $\discard \colon P \to I$ of \eqref{eq:copy_delete},\footnotemark{} which makes it into a CD category \cite{cho2019disintegration,gadducci1996algebraic,piedeleu2023introduction}.
	\footnotetext{The codomain $I$ of the deletion map is the unit of the tensor product $\otimes$ and is omitted from the string diagrams.
	Whenever $\otimes$ denotes the cartesian product of sets, it is just the singleton set $\{\ast\}$ and $\discard$ is the function mapping each element of $P$ to $\ast$.}%
	This also warrants the use of the above string diagrams.\footnotemark{}
	\footnotetext{For a target--context category with behaviors, only the left diagram in \eqref{eq:simulator} is a string diagram in the CD category, since $\eval$ is then not a morphism in this category.
	The right diagram can still be formalized, but one has to apply the shadow functor \cite[Definition 2.19]{gonda2023framework} to the simulator $s$ and interpret the diagram in the CD category $\cat{Rel}$ of sets and relations.}%

	\begin{table}[tb!]\centering
		\begin{tabular}{c|c|c|c} 
			\thead{General concept}
			& \thead{Intuitive description}
			& \thead{TM instance}
			& \thead{Spin model instance}
		\\ \hline 
			\makecell{function $C \to B$}
			& \makecell{problem to solve} 
			& \makecell{computable function}
			& \makecell{energy spectrum \\ of a spin system}
		\\ 
			\makecell{context $c \in C$}
			& \makecell{problem instance}
			& \makecell{input string}
			& \makecell{spin configuration}
		\\ 
			\makecell{behavior $b \in B$}
			& \makecell{potential answer}
			& \makecell{output string}
			& \makecell{energy}
		\\ 
			\makecell{target $t \in T$}
			& \makecell{particular solution}
			& \makecell{Turing machine}
			& \makecell{spin system}
		\\ 
			\makecell{evaluation \\ $\eval \colon T \otimes C \to B$}
			& \makecell{particular solution \\ giving answers to \\ a problem instance}
			& \makecell{running \\ Turing machines}
			& \makecell{measuring the energy \\ of spin systems}
		\\ 
			\makecell{program $p \in P$}
			& \makecell{instructions}
			& \makecell{string}
			& \makecell{parameters of \\ a spin model}
		\\ 
			\makecell{compiler \\ $s_T \colon P \to T$}
			& \makecell{collection of \\ particular solutions}
			& \makecell{e.g.\ a single \\  Turing machine}
			& \makecell{e.g.\ a spin model}
		\\ 
			\makecell{context reduction \\ $s_C \colon P \otimes C \to C$}
			& \makecell{`act of programming'}
			& \makecell{encoding programs \\ to the input tape}
			& \makecell{encoding instructions \\ to the configuration \\ of selected spins}
		\\ 
			\makecell{trivial simulator \\ $\id \colon T \otimes C \to T \otimes C$}
			& \makecell{the set of all \\ particular solutions}
			& \makecell{the set of all \\ Turing machines \\ and input strings}
			& \makecell{the set of all \\ spin systems \\ and spin configurations}
		\\ 
			\makecell{universal simulator \\ $s \colon P \otimes C \to T \otimes C$} 
			& \makecell{a universal set of \\ particular solutions}
			& \makecell{e.g.\ a universal \\ Turing machine}
			& \makecell{e.g.\ a universal \\ spin model}
		\\ 
		\end{tabular}
		\caption{Summary of framework elements.}
		\label{tab:framework}
	\end{table}
	
\section{Universality of Simulators}
\label{sec:universality}

	Among all simulators, we distinguish those that constitute a universal solution to the problems of interest.
	The simplest example, which plays a crucial role in our framework, is the so-called \newterm{trivial simulator}.
	Its programs are identical to targets and it is given by the identity map $\id \colon T \otimes C \to T \otimes C$, so that its compiler and context reduction are, respectively, 
	\begin{equation}\label{eq:trivial_sim}
		\begin{tikzpicture}
	\begin{pgfonlayer}{nodelayer}
		\node [style=morphism] (12) at (-7.75, 0) {$s_T$};
		\node [style=none] (14) at (-7.75, 1) {};
		\node [style=none] (15) at (-7.75, 1.5) {$T$};
		\node [style=none] (21) at (-7.75, -1) {};
		\node [style=none] (22) at (-7.75, -1.5) {$T$};
		\node [style=morphism] (45) at (4.625, 0) {$\;\; s_C \;\;$};
		\node [style=none] (46) at (4.625, 1) {};
		\node [style=none] (47) at (4.625, 1.5) {$C$};
		\node [style=none] (49) at (5.25, 0) {};
		\node [style=none] (50) at (4, -1) {};
		\node [style=none] (51) at (4, -1.5) {$T$};
		\node [style=none] (52) at (5.25, -1) {};
		\node [style=none] (53) at (5.25, -1.5) {$C$};
		\node [style=none] (54) at (4, 0) {};
		\node [style=none] (56) at (-5.75, 0) {$=$};
		\node [style=none] (57) at (-4, 1) {};
		\node [style=none] (58) at (-4, 1.5) {$T$};
		\node [style=none] (59) at (-4, -1) {};
		\node [style=none] (60) at (-4, -1.5) {$T$};
		\node [style=none] (61) at (7.25, 0) {$=$};
		\node [style=none] (62) at (10.75, 1) {};
		\node [style=none] (63) at (10.75, 1.5) {$C$};
		\node [style=none] (64) at (10.75, -1) {};
		\node [style=none] (65) at (10.75, -1.5) {$C$};
		\node [style=none] (66) at (9.25, -1) {};
		\node [style=none] (67) at (9.25, -1.5) {$T$};
		\node [style=bn] (68) at (9.25, 0.25) {};
		\node [style=none] (69) at (0, 0) {and};
	\end{pgfonlayer}
	\begin{pgfonlayer}{edgelayer}
		\draw (12) to (14.center);
		\draw (45) to (46.center);
		\draw (52.center) to (49.center);
		\draw (21.center) to (12);
		\draw (59.center) to (57.center);
		\draw (50.center) to (54.center);
		\draw (64.center) to (62.center);
		\draw (66.center) to (68);
	\end{pgfonlayer}
\end{tikzpicture}
	\end{equation}
	Its compiler uses all targets (as the image is $T$ itself) and its context reduction uses all contexts. 
	Intuitively, the trivial simulator is universal by `being' all possible targets in all possible contexts. 
Another simulator will be universal if it can `mimic' the behavior of the trivial simulator, i.e.\ the behavior of any target in any context. 
	
	A paradigmatic example of a universal simulator is given by a specific universal Turing machine $u$, denoted $s_u$, and described in the third column of \cref{tab:framework}. 
	In our graphical notation, it is given by 
	\begin{equation}\label{eq:univ_sim}
		\begin{tikzpicture}
	\begin{pgfonlayer}{nodelayer}
		\node [style=morphism] (12) at (-8, 0) {$s_T$};
		\node [style=none] (14) at (-8, 1.5) {};
		\node [style=none] (15) at (-8, 2) {$T$};
		\node [style=none] (21) at (-8, -1.5) {};
		\node [style=none] (22) at (-8, -2) {$P$};
		\node [style=morphism] (45) at (4.875, 0) {$\;\; s_C \;\;$};
		\node [style=none] (46) at (4.875, 1.5) {};
		\node [style=none] (47) at (4.875, 2) {$C$};
		\node [style=none] (49) at (5.5, 0) {};
		\node [style=none] (50) at (4.25, -1.5) {};
		\node [style=none] (51) at (4.25, -2) {$P$};
		\node [style=none] (52) at (5.5, -1.5) {};
		\node [style=none] (53) at (5.5, -2) {$C$};
		\node [style=none] (54) at (4.25, 0) {};
		\node [style=none] (56) at (-6, 0) {$=$};
		\node [style=none] (59) at (9.25, -1.5) {};
		\node [style=none] (60) at (9.25, -2) {$P$};
		\node [style=none] (61) at (7.25, 0) {$=$};
		\node [style=none] (62) at (9.875, 1.5) {};
		\node [style=none] (63) at (9.875, 2) {$C$};
		\node [style=none] (64) at (10.5, -1.5) {};
		\node [style=none] (65) at (10.5, -2) {$C$};
		\node [style=none] (66) at (-4.25, -1.5) {};
		\node [style=none] (67) at (-4.25, -2) {$P$};
		\node [style=bn] (68) at (-4.25, -0.75) {};
		\node [style=none] (69) at (0, 0) {and};
		\node [style=morphism] (70) at (9.875, 0) {$ \langle \ph, \ph \rangle $};
		\node [style=none] (71) at (10.5, 0) {};
		\node [style=none] (72) at (9.25, 0) {};
		\node [style=state] (73) at (-4.25, 0.5) {$u$};
		\node [style=none] (74) at (-4.25, 1.5) {};
		\node [style=none] (75) at (-4.25, 2) {$T$};
	\end{pgfonlayer}
	\begin{pgfonlayer}{edgelayer}
		\draw (12) to (14.center);
		\draw (45) to (46.center);
		\draw (52.center) to (49.center);
		\draw (21.center) to (12);
		\draw (50.center) to (54.center);
		\draw (66.center) to (68);
		\draw (59.center) to (72.center);
		\draw (64.center) to (71.center);
		\draw (70) to (62.center);
		\draw (73) to (74.center);
	\end{pgfonlayer}
\end{tikzpicture}
	\end{equation}
That is, the compiler is a constant function that maps every string in $P$ to the universal Turing machine $u \in T$.
	The context reduction is a pairing function that encodes two strings, $p \in P$ and $c \in C$, into a single string $\langle p, c \rangle \in C$.
	
	In what sense does the simulator $s_u$ simulate all particular solutions (i.e.\ all Turing machines)?
	It has the property that, for every Turing machine $t \in T$, one can find a program $p_t \in P$ such that 
	\begin{equation}\label{eq:univ_TM}
		u \bigl( \langle p_t, c \rangle \bigr) = t(c)
	\end{equation}
	for all input strings $c \in C$.
	Here, $t(c)$ denotes the behavior of the Turing machine $t$ on input $c$, which is given by the output string whenever $t$ halts on $c$.
	\Cref{eq:univ_TM} says that running the Turing machine $t$ with input $c$ is equivalent to running $u$ with input $\langle p_t, c \rangle$.
	The assignment $t \mapsto p_t$ is a specific example of a \newterm{reduction}, which is in general a map $r \colon T \to P$.
	
	Generalizing \cref{eq:univ_TM} to simulators leads to the notion of their universality.
	Namely, a simulator $s$ as in the left diagram of \eqref{eq:simulator} is \newterm{universal} if there exists a reduction $r \colon T \to P$ that makes the behavior of $s$ equivalent to that of the trivial simulator.
	In other words, the following equation
	\begin{equation}\label{eq:univ_sim_def}
		\begin{tikzpicture}
	\begin{pgfonlayer}{nodelayer}
		\node [style=none] (38) at (2.5, 1) {};
		\node [style=none] (43) at (4, 1) {};
		\node [style=none] (44) at (2.5, -1.5) {};
		\node [style=none] (45) at (2.5, -2) {$T$};
		\node [style=none] (46) at (4, -1.5) {};
		\node [style=none] (47) at (4, -2) {$C$};
		\node [style=none] (64) at (0, 0) {$=$};
		\node [style=morphism] (69) at (-3.25, 0.5) {$ \; \quad s \; \quad $};
		\node [style=none] (70) at (-4, 0.75) {};
		\node [style=none] (71) at (-2.5, 0.5) {};
		\node [style=none] (72) at (-4, 0.25) {};
		\node [style=none] (73) at (-4, -2) {$T$};
		\node [style=none] (74) at (-2.5, -1.5) {};
		\node [style=none] (75) at (-2.5, -2) {$C$};
		\node [style=morphism] (76) at (-4, -0.75) {$r$};
		\node [style=none] (77) at (-4, -1.5) {};
		\node [style=eval] (78) at (3.25, 1) {};
		\node [style=none] (79) at (3.25, 3) {};
		\node [style=none] (80) at (3.25, 3.5) {$B$};
		\node [style=none] (81) at (-4, 2) {};
		\node [style=none] (82) at (-2.5, 2) {};
		\node [style=eval] (83) at (-3.25, 2) {};
		\node [style=none] (84) at (-3.25, 3) {};
		\node [style=none] (85) at (-3.25, 3.5) {$B$};
	\end{pgfonlayer}
	\begin{pgfonlayer}{edgelayer}
		\draw (46.center) to (43.center);
		\draw (44.center) to (38.center);
		\draw (74.center) to (71.center);
		\draw (72.center) to (70.center);
		\draw (76) to (72.center);
		\draw (77.center) to (76);
		\draw (78) to (79.center);
		\draw (83) to (84.center);
		\draw (70.center) to (81.center);
		\draw (71.center) to (82.center);
	\end{pgfonlayer}
\end{tikzpicture}
	\end{equation}
	must be satisfied.
	In the case of $s_u$, \cref{eq:univ_sim_def} amounts to \cref{eq:univ_TM} \cite[Example 3.6]{gonda2023framework}.
	
	Universal simulators, as introduced in our framework, satisfy a more complicated requirement than presented here. 
	For example, equality in \eqref{eq:univ_sim_def} is weakened to an inequality, which intuitively says that the behavior of the (reduced) universal simulator need not equal that of the trivial one, but it may subsume it in a suitable way.
	This gives the framework more versatility and expressive power. 
	
	This generality is needed for spin model universality. 
	Let us briefly recount this example, whose elements are also summarized in \cref{tab:framework}. 
	The targets consist of spin systems, while programs parametrize a subset of them via the compiler $s_T$. 
	The image of the compiler $s_T$ is the set of spin systems whose universality we want to  investigate. 
	A spin model is a collection of spin systems. 
	
	Each spin system consists of interacting degrees of freedom called spins.
	The so-called spin configuration, modeled as a context $c \in C$, specifies a value of each of the spins.
	Much of the relevant information about a spin system can be obtained from studying its energy values.
	This is captured by the evaluation function $\eval$, which maps each spin system $t \in T$ together with a matching spin configuration $c \in C$ to its energy. 
	The reduction $r$ maps the spin system to be simulated, an arbitrary $t \in T$, to a parameter $p \in P$, which may encode the size of the spin system, the values of the coupling strengths and the number of levels of the spins.  
	The spin system that should simulate $t$ is given by applying the compiler to obtain $s_T(r(t)) \in T$. 
	Finally, the context reduction maps $p$ together with the spin configuration $c$ (of the spin system $t$) to a configuration $s_C(p,c) \in C$ of the spin system $s_T(p)$.

	The full story of spin model universality is more involved, see \cite[Section 3.2]{gonda2023framework}.
	For example, while above we use the terminology of functions for morphisms (such as the context reduction), they are in fact relations, i.e.\ partial multi-valued functions.
	Moreover, the number of degrees of freedom of the output of a morphism can only grow polynomially with the input number of degrees of freedom.
	This makes the proven universality stronger as it means, for example, that the reduction cannot map $t$ to an arbitrarily large $r(t)$.
	
	Other examples of universality in mathematics can be expressed as universal simulators, such as completeness for a computational complexity class (such as NP), dense subset of a topological space, cofinal subset of a preorder, or a universal Borel set \cite{gonda2023framework}.
	We expect that more will follow, including universal graphs \cite{Ra64}, universal neural networks \cite{Cy89,Ho91b,Cs01,Le07}, universal grammars \cite{Ch65}, or universal quantum spin models \cite{Cu17,Zh21}.
	
	Examples of universality in fields further away from mathematics\,---\,such as programmable synthesis for chemical \cite{gromski2020universal}, biological \cite{blackiston2023biological}, and other complex systems \cite{doursat2012morphogenetic}\,---\,take more work to cast within our framework, but we may also learn more about them by doing so.
	For instance, Blackiston et al.\ \cite{blackiston2023biological} mention that ``Xenobots \cite{blackiston2021cellular} are a good example in which to examine the ... blurring of lines between classical robots and organisms... They challenge us to expand traditional notions of a `program' beyond linear code written by a human to probabilistic, parallel, naturally evolving control policies...''.

\section{No-go Theorem for Universality}
\label{sec:no-go}
	
	In the rest of the article, we focus on results derived within our framework.
	In this section, we describe a theorem providing conditions under which a simulator cannot be universal \cite[Theorem 3.19]{gonda2023framework}.
	
	To illustrate it, let us show by contradiction that a universal spin model must contain infinitely many spin systems.
	The no-go theorem generalizes this argument to any instance of our framework. 
	
	To this end, consider an arbitrary spin model specified by a compiler $s_T \colon P \to T$ as in \cref{sec:universality}.
	Moreover, we assume that it is a finite spin model, i.e.\ that $P$ is a finite set.
	The compiler gives a subset of all spin systems that are `used' by the simulator via its image
	\begin{equation}\label{eq:compiler_image}
		\mathrm{Im}(s_T) \coloneqq \Set{ s_T (p)  \given  p\in P }.
	\end{equation}
	There is a maximum number of energy levels (call it $S_{\rm max}$) 
	these spin systems can display.
	This is due to the fact that each spin system has a finite number of spin configurations and thus a finite spectrum (i.e.\ the range of attainable energy values).
	
	In order for such a simulator to be universal, it needs to be able to reproduce the behavior of every other spin system. 
	However, due to the notion of evaluation (and the preorder replacing equality relation in \eqref{eq:univ_sim_def} \cite[Section 3.2]{gonda2023framework}), a given spin system can only reproduce behaviors of those with no larger spectrum. 
	It follows that any spin system whose spectrum size is larger than $S_{\rm max}$ cannot have its behavior reproduced by the above finite spin model.
	Since one can construct spin systems with arbitrarily large spectra, the proposed simulator cannot be universal.
	
	In the general case, we replace (i) the idea of reproducing the behavior of a spin system by the `lax context-reduction preorder' among targets, (ii) the size of a spectrum by a function that is order-preserving relative to it, and (iii) the unboundedness of spectra sizes by the supremum of this function over all elements of $T$.
	This yields a theorem to rule out universality for a given simulator, in analogy to ruling out universality of finite spin models. 

\section{Parsimony of Universal Simulators}\label{sec:parsimony}
	
	The compiler image of a spin model must be infinite if the spin model is to be universal.
	On the other hand, the simulator of Turing machines $s_u$ from \eqref{eq:univ_sim} uses a single target\,---\,a universal Turing machine\,---\,to simulate all others. 
	Its compiler image is a singleton set, which is why we refer to it as a \textbf{singleton simulator}.
	
	Universal simulators thus exhibit major differences in how they make use of the space of targets.
	The singleton simulator $s_u$ uses only one target.
	On the other extreme is the trivial simulator $\id$ from \eqref{eq:trivial_sim}, which uses all targets.

	We formalize these distinctions in terms of a \textbf{parsimony preorder} among universal simulators.
	Among the above examples, $s_u$ should be highly parsimonious, while $\id$ should be not parsimonious at all.
	The idea of the ordering is the following: 
	A non-parsimonious simulator, such as $\id$, can be compressed to obtain a more parsimonious one, such as $s_u$.
	On the other hand, a simulator that is already `resource-saving' cannot be compressed to generate one that uses all targets.
	
	The notion of compression of simulators consists of a post-processing $q$ and a pre-processing $r$ giving rise to the overall mapping
	\begin{equation}\label{eq:morphism}
		\begin{tikzpicture}
	\begin{pgfonlayer}{nodelayer}
		\node [style=none] (4) at (5.5, 1.75) {};
		\node [style=none] (5) at (5, 3.25) {};
		\node [style=none] (6) at (3.25, 3.25) {};
		\node [style=none] (12) at (0, 0) {$\mapsto$};
		\node [style=none] (33) at (2.5, 1.75) {};
		\node [style=morphism] (2) at (4, 1.75) {$\qquad q \qquad $};
		\node [style=none] (34) at (4, 1.75) {};
		\node [style=none] (35) at (5, 1.75) {};
		\node [style=none] (36) at (3.25, 1.75) {};
		\node [style=morphism] (37) at (-3.5, 0) {$ \quad s \quad $};
		\node [style=none] (38) at (-4.25, 0) {};
		\node [style=none] (39) at (-4.25, 1.5) {};
		\node [style=none] (40) at (-4.25, 2) {$T$};
		\node [style=none] (41) at (-2.75, 1.5) {};
		\node [style=none] (42) at (-2.75, 2) {$C$};
		\node [style=none] (43) at (-2.75, 0) {};
		\node [style=none] (44) at (-4.25, -1.5) {};
		\node [style=none] (45) at (-4.25, -2) {$P$};
		\node [style=none] (46) at (-2.75, -1.5) {};
		\node [style=none] (47) at (-2.75, -2) {$C$};
		\node [style=none] (48) at (3.25, 3.75) {$T$};
		\node [style=none] (49) at (5, 3.75) {$C$};
		\node [style=morphism] (50) at (4.75, 0.25) {$ \quad s \quad $};
		\node [style=none] (51) at (4, 0.5) {};
		\node [style=none] (52) at (5.5, 0.25) {};
		\node [style=none] (53) at (4, 0) {};
		\node [style=none] (54) at (3.25, -3.75) {$P'$};
		\node [style=none] (55) at (5.5, -3.25) {};
		\node [style=none] (56) at (5.5, -3.75) {$C$};
		\node [style=morphism] (57) at (4, -1.25) {$r$};
		\node [style=bn] (58) at (3.25, -2.5) {};
		\node [style=none] (59) at (2.5, -1.25) {};
		\node [style=none] (60) at (3.25, -3.25) {};
	\end{pgfonlayer}
	\begin{pgfonlayer}{edgelayer}
		\draw (36.center) to (6.center);
		\draw (35.center) to (5.center);
		\draw (46.center) to (43.center);
		\draw (43.center) to (41.center);
		\draw (44.center) to (38.center);
		\draw (38.center) to (39.center);
		\draw (55.center) to (52.center);
		\draw (53.center) to (51.center);
		\draw (52.center) to (4.center);
		\draw [in=-90, out=90] (51.center) to (34.center);
		\draw [in=-90, out=150] (58) to (59.center);
		\draw (59.center) to (33.center);
		\draw [bend left] (57) to (58);
		\draw (57) to (53.center);
		\draw (60.center) to (58);
	\end{pgfonlayer}
\end{tikzpicture}
	\end{equation}
	called a \textbf{simulator morphism} and denoted by $(q,r) \colon s \to s'$ where $s'$ is the simulator on the right-hand side.
	Additional conditions not spelled out here are placed on $q$ and $r$ \cite[Section 4]{gonda2023framework}.
	As a consequence, morphisms can be composed and give rise to a category of simulators and their morphisms.
	
A universal simulator $s'$ is at least as parsimonious as $s$ if there exists such a simulator morphism mapping $s$ to $s'$.
	Based on this definition, we prove in \cite{gonda2023framework} that the singleton simulator $s_u$ is strictly more parsimonious than the trivial simulator.
	This means that there exists a simulator morphism of type $\id \to s_u$, but no such morphism of type $s_u \to \id$.
	
	Our proof is based on an abstract argument that can be applied more generally.
	Specifically, \cite[Theorem 4.10]{gonda2023framework} provides conditions that guarantee a simulator to be at least as parsimonious as the trivial simulator.
	On the other hand, \cite[Theorem 4.13]{gonda2023framework} gives necessary conditions for the existence of a simulator morphism from a generic simulator to the trivial one.
	
	Note that since the trivial simulator always exists and is universal, every instance of our framework features a universal simulator. 
	A more nuanced property is whether there exists a `non-trivial' universal simulator\,---\,one that is strictly more parsimonious than the trivial simulator.
	Universality of Turing machines is non-trivial in this sense.
Whether the 2D Ising spin model (or other universal spin models) are non-trivial universal simulators is yet to be proved in this framework. 

\section{Unreachability}\label{sec:unreachability}

For Turing machines, one can use the diagonal argument (i.e.\ self-reference and negation) to show that not all functions are computable.
	For example, the halting problem can be proven to be uncomputable via this reasoning. 
	In the Turing machine instance of our framework, this says that the set of all functions that are `reachable by evaluating a target', i.e.\ satisfy
	\begin{equation}\label{eq:reachable}
		\begin{tikzpicture}
	\begin{pgfonlayer}{nodelayer}
		\node [style=none] (70) at (2, 0.25) {};
		\node [style=none] (71) at (3.5, 0) {};
		\node [style=none] (72) at (2, -0.25) {};
		\node [style=none] (74) at (3.5, -1.5) {};
		\node [style=none] (75) at (3.5, -2) {$C$};
		\node [style=none] (81) at (2, 0.5) {};
		\node [style=none] (82) at (3.5, 0.5) {};
		\node [style=eval] (83) at (2.75, 0.5) {};
		\node [style=none] (84) at (2.75, 1.5) {};
		\node [style=none] (85) at (2.75, 2) {$B$};
		\node [style=state] (86) at (2, -1) {$t$};
		\node [style=none] (87) at (0, 0) {$=$};
		\node [style=morphism] (88) at (-2.5, 0) {$f$};
		\node [style=none] (89) at (-2.5, 1.5) {};
		\node [style=none] (90) at (-2.5, 2) {$B$};
		\node [style=none] (91) at (-2.5, -1.5) {};
		\node [style=none] (92) at (-2.5, -2) {$C$};
	\end{pgfonlayer}
	\begin{pgfonlayer}{edgelayer}
		\draw (74.center) to (71.center);
		\draw (72.center) to (70.center);
		\draw (83) to (84.center);
		\draw (70.center) to (81.center);
		\draw (71.center) to (82.center);
		\draw (86) to (72.center);
		\draw (91.center) to (88);
		\draw (88) to (89.center);
	\end{pgfonlayer}
\end{tikzpicture}
	\end{equation}
	for some Turing machine $t$, is restricted to computable functions.
	Whenever there are morphisms of type $C \to B$ that cannot be expressed in such a way, we say that the trivial simulator has \newterm{unreachability}.
	This notion generalizes the idea of uncomputability and undecidability.
	
	Moreover, the categorical version of the diagonal argument due to Lawvere \cite{La69b} (and extended by Roberts \cite{Ro21}) can be generalized to our setting \cite[Theorem 5.4]{gonda2023framework}.
	In simple terms, it says that if the space of targets is isomorphic to that of contexts (i.e.\ $T \cong C$) and there is a morphism without a fixed point, then the trivial simulator has unreachability.
In the long paper, we illustrate the use of this fixed point theorem by recovering the well-known fact that a universal Turing machine must be non-halting on some inputs.
	
\begin{figure}[t]\centering
		\includegraphics[width=.9\columnwidth]{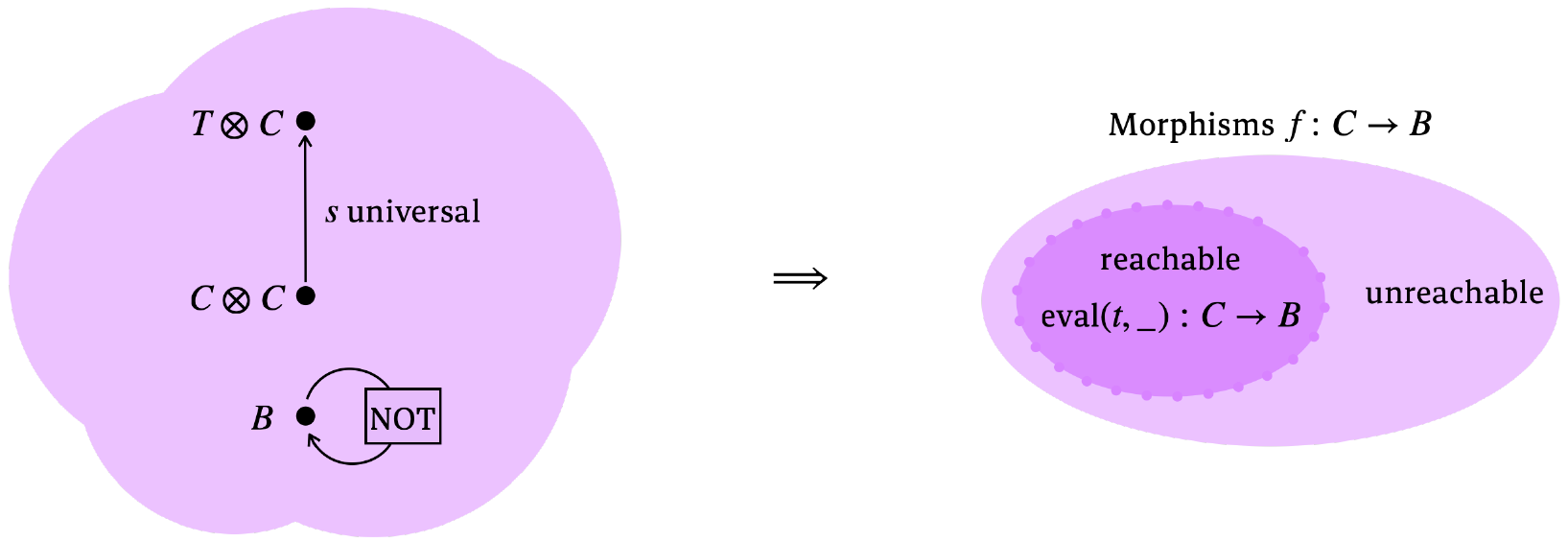}
		\caption{A universal simulator with $P \cong C$ together with a fixed-point-free morphism $B\to B$ entails unreachability.}
		\label{fig:unreachability}
	\end{figure}

	One of the drawbacks of the fixed point theorem is that it presupposes a direct self-reference via the isomorphism between targets and contexts.
	To apply it to instances where $T$ is not isomorphic to $C$, we need to invoke universality.
	In particular, a universal simulator whose programs $P$ are isomorphic to contexts can introduce self-reference indirectly.
	As we show in \cite[Theorem 5.10]{gonda2023framework} and depict in \cref{fig:unreachability}, one can guarantee unreachability of the trivial simulator if there is a universal simulator of type $C \otimes C \to T \otimes C$ and a morphism of type $B \to B$ without a fixed point.
		
	Besides unreachability, this result can be used as another no-go theorem for universality.
	Namely, a contrapositive of the statement says: If the trivial simulator does not have unreachability and there is a fixed-point-free morphism $B \to B$, then there is no universal simulator whose programs are isomorphic to contexts.
	We illustrate this application in \cite[Example 5.11]{gonda2023framework} by considering the ambient category of sets and functions.
	To this end, we consider $C$ to be an arbitrary set, $B$ the two-element set, and $T$ the power set of $C$ which can be described as the set of all functions of type $C \to B$.
	If the evaluation map is given as usual by $\eval(t,c) = t(c)$, then the trivial simulator does not have unreachability in this instance of our framework. 
	Furthermore, the negation that swaps elements of $B$ is a fixed-point-free morphism on behaviors.
	Thus, there can be no universal simulator whose programs are given by $C$.
	This fact is equivalent to the conclusion of Cantor's Theorem, which says that there is no surjection of type $C \to B^C$.
	Similar arguments can be applied to closed monoidal categories more generally.

	\section*{Acknowledgements}
	The framework on which this article is based was develop in collaboration with Tobias Reinhart and Sebastian Stengele.
	This research was funded by the Austrian Science Fund (FWF) via the START Prize Y1261-N.
	For open access purposes, the authors have applied a CC BY public copyright license to any accepted manuscript version arising from this submission.
	

\begin{thebibliography}{23}
\providecommand{\natexlab}[1]{#1}
\providecommand{\url}[1]{\texttt{#1}}
\expandafter\ifx\csname urlstyle\endcsname\relax
  \providecommand{\doi}[1]{doi: #1}\else
  \providecommand{\doi}{doi: \begingroup \urlstyle{rm}\Url}\fi

\bibitem[Blackiston et~al.(2021)Blackiston, Lederer, Kriegman, Garnier,
  Bongard, and Levin]{blackiston2021cellular}
D.~Blackiston, E.~Lederer, S.~Kriegman, S.~Garnier, J.~Bongard, and M.~Levin.
\newblock A cellular platform for the development of synthetic living machines.
\newblock \emph{Science Robotics}, 6\penalty0 (52):\penalty0 eabf1571, 2021.

\bibitem[Blackiston et~al.(2023)Blackiston, Kriegman, Bongard, and
  Levin]{blackiston2023biological}
D.~Blackiston, S.~Kriegman, J.~Bongard, and M.~Levin.
\newblock Biological robots: Perspectives on an emerging interdisciplinary
  field.
\newblock \emph{Soft Robotics}, 10\penalty0 (4):\penalty0 674--686, 2023.
\newblock \doi{10.1089/soro.2022.0142}.

\bibitem[Cho and Jacobs(2019)]{cho2019disintegration}
K.~Cho and B.~Jacobs.
\newblock Disintegration and {B}ayesian inversion via string diagrams.
\newblock \emph{Mathematical Structures in Computer Science}, 29\penalty0
  (7):\penalty0 938--971, 2019.
\newblock \doi{10.1017/s0960129518000488}.

\bibitem[Chomsky(1965)]{Ch65}
N.~Chomsky.
\newblock \emph{{Aspects of the theory of syntax}}.
\newblock MIT Press, 1965.
\newblock
  \href{https://mitpress.mit.edu/books/aspects-theory-syntax}{https://mitpress.mit.edu/books/aspects-theory-syntax}.

\bibitem[Csaji(2001)]{Cs01}
B.~C. Csaji.
\newblock \emph{{Approximation with Artificial Neural Networks}}.
\newblock Master Thesis, Eindhoven University of Technology, 2001.

\bibitem[Cubitt et~al.(2018)Cubitt, Montanaro, and Piddock]{Cu17}
T.~S. Cubitt, A.~Montanaro, and S.~Piddock.
\newblock {Universal quantum Hamiltonians}.
\newblock \emph{Proc. Natl. Acad. Sci.}, 38:\penalty0 9497--9502, 2018.
\newblock \doi{10.1073/pnas.1804949115}.

\bibitem[Cybenko(1989)]{Cy89}
G.~Cybenko.
\newblock {Approximation by superpositions of a sigmoidal function}.
\newblock \emph{MCSS}, 4:\penalty0 455, 1989.
\newblock \doi{10.1007/BF02551274}.

\bibitem[{De las Cuevas} and Cubitt(2016)]{De16b}
G.~{De las Cuevas} and T.~S. Cubitt.
\newblock {Simple universal models capture all classical spin physics}.
\newblock \emph{Science}, 351:\penalty0 1180--1183, 2016.
\newblock \doi{10.1126/science.aab3326}.

\bibitem[Doursat et~al.(2012)Doursat, Sayama, and
  Michel]{doursat2012morphogenetic}
R.~Doursat, H.~Sayama, and O.~Michel.
\newblock \emph{Morphogenetic engineering: Toward programmable complex
  systems}.
\newblock Springer, 2012.
\newblock \doi{10.1007/978-3-642-33902-8}.

\bibitem[Gadducci(1996)]{gadducci1996algebraic}
F.~Gadducci.
\newblock On the algebraic approach to concurrent term rewriting.
\newblock \emph{Bulletin-European Association for Theoretical Computer
  Science}, 59:\penalty0 412--413, 1996.

\bibitem[Gonda et~al.(2024)Gonda, Reinhart, Stengele, and
  Coves]{gonda2023framework}
T.~Gonda, T.~Reinhart, S.~Stengele, and G.~D.~l. Coves.
\newblock A framework for universality in physics, computer science, and
  beyond.
\newblock \emph{Compositionality}, 6\penalty0 (3), 2024.
\newblock \doi{10.32408/compositionality-6-3}.

\bibitem[Gromski et~al.(2020)Gromski, Granda, and Cronin]{gromski2020universal}
P.~S. Gromski, J.~M. Granda, and L.~Cronin.
\newblock Universal chemical synthesis and discovery with ‘the chemputer’.
\newblock \emph{Trends in Chemistry}, 2\penalty0 (1):\penalty0 4--12, 2020.

\bibitem[Hornik(1991)]{Ho91b}
K.~Hornik.
\newblock {Approximation capabilities of multilayer feedforward networks}.
\newblock \emph{Neural Netw.}, 4:\penalty0 251, 1991.
\newblock \doi{10.1016/0893-6080(91)90009-T}.

\bibitem[Lawvere(1969)]{La69b}
F.~W. Lawvere.
\newblock {Diagonal arguments and cartesian closed categories}.
\newblock \emph{Lecture Notes in Mathematics}, 92:\penalty0 134--145, 1969.
\newblock \doi{10.1007/bfb0080769}.

\bibitem[{Le Roux} and Bengio(2008)]{Le07}
N.~{Le Roux} and Y.~Bengio.
\newblock Representational power of restricted {B}oltzmann machines and deep
  belief networks.
\newblock \emph{Neural Comput.}, 20:\penalty0 1631, 2008.
\newblock \doi{10.1162/neco.2008.04-07-510}.

\bibitem[Li and Vit{\'a}nyi(2008)]{Li2008}
M.~Li and P.~Vit{\'a}nyi.
\newblock \emph{An introduction to Kolmogorov complexity and its applications},
  volume~3.
\newblock Springer, 2008.

\bibitem[Piedeleu and Zanasi(2023)]{piedeleu2023introduction}
R.~Piedeleu and F.~Zanasi.
\newblock An introduction to string diagrams for computer scientists.
\newblock 2023.
\newblock \href{https://arxiv.org/abs/2305.08768}{arXiv:2305.08768}.

\bibitem[Rado(1964)]{Ra64}
R.~Rado.
\newblock {Universal graphs and universal functions}.
\newblock \emph{Acta Arithmetica}, 9:\penalty0 331--340, 1964.
\newblock \doi{10.4064/aa-9-4-331-340}.

\bibitem[Reinhart et~al.(In preparation)Reinhart, Engel, and {De les
  Coves}]{Re23}
T.~Reinhart, B.~Engel, and G.~{De les Coves}.
\newblock A modular framework for simulations in classical spin models:
  {U}niversality and its consequences.
\newblock In preparation.

\bibitem[Roberts(2023)]{Ro21}
D.~M. Roberts.
\newblock Substructural fixed-point theorems and the diagonal argument: {T}heme
  and variations.
\newblock \emph{Compositionality: the open-access journal for the mathematics
  of composition}, 5, 2023.
\newblock \doi{10.32408/compositionality-5-8}.

\bibitem[Sol{\'{e}} and Goodwin(2000)]{So00}
R.~V. Sol{\'{e}} and B.~Goodwin.
\newblock \emph{{Signs of Life: How complexity pervades biology}}.
\newblock Basic Books, 2000.

\bibitem[Thurner et~al.(2018)Thurner, Hanel, and Klimek]{Th18}
S.~Thurner, R.~Hanel, and P.~Klimek.
\newblock \emph{{Introduction to the Theory of Complex Systems}}.
\newblock Oxford University Press, 2018.

\bibitem[Zhou and Aharonov(2021)]{Zh21}
L.~Zhou and D.~Aharonov.
\newblock Strongly universal {H}amiltonian simulators.
\newblock 2021.
\newblock \doi{10.48550/arXiv.2102.02991}.

\end{thebibliography}

\end{document}